\documentclass[a4paper,11pt]{article}
\usepackage{graphicx,amssymb,bm,latexsym,color,epsf}
\makeatletter
\@addtoreset{equation}{section}

\makeatother
\newcommand{\be}{\begin{equation}}
\newcommand{\ee}{\end{equation}}
\newcommand{\bea}{\begin{eqnarray}}
\newcommand{\eea}{\end{eqnarray}}
\newcommand{\vs}[1]{\vspace{#1 mm}}
\newcommand{\hs}[1]{\hspace{#1 mm}}
\renewcommand{\a}{\alpha}
\renewcommand{\b}{\beta}
\renewcommand{\c}{\gamma}
\newcommand{\G}{\Gamma}
\renewcommand{\d}{\delta}
\newcommand{\e}{\epsilon}
\newcommand{\s}{\sigma}
\renewcommand{\t}{\theta}
\newcommand{\vp}{\varphi}
\newcommand{\la}{\lambda}
\newcommand{\pa}{\partial}
\newcommand{\nn}{\nonumber\\}
\newcommand{\p}[1]{(\ref{#1})}
\newcommand{\lan}{\langle}
\newcommand{\ran}{\rangle}

\newcommand{\br}{\bar R}
\newcommand{\bg}{\bar g}

\newcommand{\bnabla}{\bar\nabla}

\newcommand{\Det}{{\rm Det}}
\newcommand{\Tr}{{\rm Tr}}
\newcommand{\ta}{\tilde \a}
\newcommand{\tb}{\tilde \b}
\newcommand{\tc}{\tilde \c}

\begin{document}

\begin{titlepage}

\renewcommand{\thefootnote}{\fnsymbol{footnote}}
\begin{flushright}
KU-TP 069 \\
\today
\end{flushright}

\vs{10}
\begin{center}
{\Large\bf Background Scale Independence in Quantum Gravity}
\vs{15}

{\large
Nobuyoshi Ohta\footnote{e-mail address: ohtan@phys.kindai.ac.jp}
} \\
\vs{10}

{\em Department of Physics, Kindai University,
Higashi-Osaka, Osaka 577-8502, Japan
\\
and
\\
 Maskawa Institute for Science and Culture,
Kyoto Sangyo University, Kyoto 603-8555, Japan}

\vs{15}
{\bf Abstract}
\end{center}

We study the background scale independence in single-metric approximation to
the functional renormalization group equation (FRGE) for quantum gravity
and show that it is possible to formulate it without using higher-derivative gauge fixing
in arbitrary dimensions if we adopt the Landau gauge and suitable cutoff scheme.
We discuss this problem for both the linear and exponential splits of the metric into
background and fluctuations. The obtained modified Ward identity for the global rescalings
of the background metric can be combined with the FRGE
to give a manifestly scale-invariant solution.
An explicit example of the FRGE is given
for four-dimensional $f(R)$ gravity in this framework.


\end{titlepage}
\newpage
\setcounter{page}{2}
\renewcommand{\thefootnote}{\arabic{footnote}}
\setcounter{footnote}{0}

\section{Introduction}

Asymptotic safety is one of the promising approaches to the formulation of quantum gravity
within the framework of conventional field theory~\cite{W}.
This program considers the possibility of having an interacting quantum field theory of gravitation,
originated from a non-Gaussian ultraviolet (UV) fixed point in the theory space,
in the nonperturbative renormalization group (RG) framework~\cite{wetterich,Morris0,Reuter}.
The existence of such a fixed point (scaling solution) would permit having an RG trajectory in the theory space
flowing to it and characterized by all the dimensionless couplings remaining finite when the UV cutoff is removed.
If the number of relevant operators is finite, this theory has predictive power.
The approach has produced a wealth of results. For reviews and introductions, see~\cite{RS,Percacci,Litim}.

To actually pursue this line, one has to make approximation such as truncation,
derivative expansion etc. Most approaches keep a finite number of local operators in the effective
action. These polynomial truncations may be viewed as based on a small curvature expansion
and have arrived at amazingly higher-order expansion up to 34th order in the scalar curvature~\cite{fallslitim}.
However, this still might not be good enough to draw the conclusion that asymptotic safety is
achieved convincingly as long as one retains a finite number of operators.
To go beyond this, one has to keep an infinite number of operators, which enables one to treat them
without expansion. In the so-called $f(R)$ approximation~\cite{cpr1}-\cite{FO}, a Lagrangian of
the form $f(R)$, not just a polynomial expansion, is used in which all possible forms in the scalar
curvature are considered.

In this approach, a momentum cutoff $k$ is introduced. The quantum effective action is
recovered when we take the $k\to 0$ limit.
It has been pointed out that a problem arises in a large-curvature regime
where $k$ is smaller than the minimum eigenvalues of the Laplacian~\cite{dsz1,opv2,FO}.
If the spectrum of all operators has a finite gap $\delta$, then for $k<\delta$
the flow equation does not integrate out any modes.
This raises the question of the meaning of coarse-graining on length scales
that are larger than the size of the manifold.

Recently, Morris has pointed out that this problem arises because the background independence is
not respected in the single-metric formulation adapted in these approaches~\cite{Morris}.
Background independence means that physics should not depend on the choice of the background.
There are several approaches to address this issue. One is to use bimetric
truncations~\cite{manrique,beckerreuter} and require shift invariance.
Another is to solve the modified Ward identity (mWI) for the shift transformation and
the flow equation simultaneously~\cite{dm3,lms,dms}.

Shift invariance is a symmetry that keeps the classical action invariant.
If one uses a linear split of the metric into background $\bg$ and fluctuation $h$:
\bea
g_{\mu\nu} = \bg_{\mu\nu} + h_{\mu\nu},
\label{linear}
\eea
the classical action is invariant under a simultaneous shift of the background and fluctuation:
\bea
\d \bg_{\mu\nu} = \e_{\mu\nu}, \qquad
\d h_{\mu\nu} = -\e_{\mu\nu}.
\eea
This should be a symmetry of the effective action, but is broken by gauge fixing and cutoff terms
in the process of quantization and regularization.
The problem associated with this was pointed out in \cite{LP}.

Morris~\cite{Morris} has restricted the shift transformation to constant rescaling
as a first step towards realizing the full invariance under shift symmetry
and discussed the above problem associated with the coarse-graining. 
The basic idea is that it is wrong to relate the cutoff scale $k$ to the fixed background;
rather all background metrics of different overall scales should be treated equally.
He has been able to derive the mWI corresponding to the rescaling and show that it is compatible
with the functional renormalization group equation (FRGE). 
It is found that the resulting solution to the FRGE is written in terms of scale-invariant variables.
The background metric is no longer just a fixed one but is replaced by a dynamical variable,
and we have to consider that it describes ensemble of metrics of different scales.
In this way theories on a continuous infinity of manifolds are incorporated
and the above problem related to the coarse-graining may be resolved.
Unfortunately, he was able to do this only for 
six dimensions.

More recently it has been pointed out that the formulation can be extended to arbitrary
dimensions if one uses 1) the exponential split of the metric into background and fluctuation,
2) higher-derivative gauge fixing and 3) special cutoff scheme~\cite{PV}.
The exponential split of the metric is defined by~\cite{KN}
\bea
g_{\mu\nu} = \bg_{\mu\rho}(e^h)^\rho_{\nu}.
\label{exp}
\eea
It would be an interesting problem to study if the above three conditions are absolutely necessary.

In this paper we point out that Morris's formulation with the usual gauge fixing may be
extended to arbitrary dimensions if one uses the Landau gauge and a cutoff scheme similar to
but slightly different from that of Percacci and Vacca~\cite{PV}.
We show why the Landau gauge is singled out in order to make this formulation work
and emphasize that it is not necessary to adopt exponential split. 
Since the exponential parametrization~\p{exp} has been shown to have various virtues compared to
the linear split~\cite{opv,opv2,FO,nink,GKL,opp}, even though the consistent mWI can be derived
in arbitrary dimensions with the linear split, it is interesting to see if the result may also be
extended to the exponential split in arbitrary dimensions.
We thus discuss this approach in both linear and exponential splits of the metric.

This paper is organized as follows.
In the next section, we set off to discuss how to implement the rescaling invariance
with the linear split in arbitrary dimensions. We first define the global rescaling transformation and
show how the invariance of the gauge-fixing term forces us to choose the Landau gauge
if we use the ordinary gauge-fixing condition. This is also noted by Morris, but the use of
the auxiliary field makes the discussion more transparent. Then, motivated by Percacci and Vacca~\cite{PV},
we adopt what is called a ``pure'' cutoff scheme~\cite{NP} that breaks the invariance but in a way
that is compatible with the FRGE in arbitrary dimensions.
In this process we have to modify the ghost part slightly compared to \cite{PV}
because our gauge fixing is the usual one without higher derivatives.
In this way we derive the mWI and arrive at a solution that is compatible with scale invariance.
In sect.~3, we go on to discuss the same problem with the exponential split.
We obtain basically the same result with usual gauge fixing but with a suitably modified pure cutoff
in arbitrary dimensions.
In sect.~4, we discuss how the FRGE looks with our pure cutoff, but the result is rather complicated.
We leave discussions of the solution for future study.
Section~5 is devoted to conclusions.

\section{Modified Ward identity in the linear split}

In this section, we derive the mWI in the linear split~\p{linear} valid for arbitrary dimensions,
and show its compatibility with FRGE.

In either parametrization~\p{linear} or \p{exp}, the partition function is given by
\bea
Z[g_{\mu\nu},J^{\a\b}] = \int[{\cal D} h_{\mu\nu}] \exp\left[ -S_0[g_{\mu\nu},h_{\mu\nu}]
+\int J^{\mu\nu} h_{\mu\nu} \right].
\eea
Throughout this paper, the indices are raised or lowered by the background metric $\bg$ unless otherwise stated.
We consider $d$-dimensional Euclidean spacetime.
By $\int$, we mean $\int d^d x$ and the $\sqrt{\bg}$ factor is absorbed into the definition of $J^{\mu\nu}$.

\subsection{Rescaling transformation}

We would like to impose background scale independence under the constant rescaling
\bea
\d\bg_{\mu\nu} = 2\e \bg_{\mu\nu},
\label{t1}
\eea
together with
\bea
\d h_{\mu\nu} = -2\e \bg_{\mu\nu},
\label{t2}
\eea
so that the total metric~\p{linear} does not change.
We decompose the fluctuation field into traceless and trace parts:
\bea
h_{\mu\nu} = h_{\mu\nu}^T + \frac{1}{d} \bg_{\mu\nu} h.
\label{dec}
\eea
The variation gives
\bea
\d h_{\mu\nu}=\d h_{\mu\nu}^T+\frac{1}{d}\d\bg_{\mu\nu}h +\frac{1}{d}\bg_{\mu\nu} \d h.
\label{t3}
\eea
Using \p{t1} and \p{t2}, and taking the trace, we find
\bea
\d h =-2\e(h+d).
\label{t4}
\eea
Substituting \p{t4} back into \p{t3}, we get
\bea
\d h_{\mu\nu}^T = 0.
\label{t5}
\eea

The gauge-fixing and Faddeev-Popov (FP) terms may be obtained most easily by using the BRST transformation.
This can be obtained by replacing the parameters of the coordinate transformation by the ghost:
\bea
\d_B g_{\mu\nu} = -\d\la\, ( g_{\mu\a}\nabla_\nu C^\a + g_{\nu\a}\nabla_\mu C^\a),
\eea
where $\d\la$ is an anticommuting parameter.
This should be regarded as the transformation of the fluctuation field $h_{\mu\nu}$ since the quantum gauge
transformation is generated when $\bg_{\mu\nu}$ is held fixed.
\bea
\d_B h_{\mu\nu} = -\d\la\, ( g_{\mu\a}\nabla_\nu C^\a + g_{\nu\a}\nabla_\mu C^\a),
\label{brst1}
\eea
We should note that here the covariant derivative is defined by the whole metric.
The BRST transformation for other fields is derived by the requirement of the nilpotency of
the transformation:
\bea
\d_B C^\mu \hs{-2}&=&\hs{-2} \d\la C^\rho \pa_\rho C^\mu,~~~
\d_B \bar C_\mu = i \d\la\, B_\mu, ~~~
\d_B B_\mu = 0,
\label{brst2}
\eea
where $\bar C_\mu$ is the FP anti-ghost and $B_\mu$ is an auxiliary field that enforces
the gauge-fixing condition.
The gauge-fixing function is
\bea
F_\mu = \bg^{\rho\nu} \Big(\bnabla_\rho h_{\mu\nu}-\frac{b+1}{d} \bnabla_\mu h_{\rho\nu}\Big),
\label{gff}
\eea
where $b$ is a gauge parameter. Here and in what follows, the bar on the covariant derivative
means that it is constructed with the background metric.
Note that $\bnabla_\mu$ is invariant under the rescaling transformation.
The gauge-fixing and FP terms are then given as~\cite{opp}
\bea
{\cal L}_{GF+FP}/\sqrt{\bg} \hs{-2}&=&\hs{-2}
i \d_B \left[\bg^{\mu\nu} \bar C_\mu \left(F_\nu+\frac{a}{2} B_\nu\right)\right]/\d\la \nn
\hs{-2} &=&\hs{-2} - \bg^{\mu\nu} B_\mu \left( F_\nu +\frac{a}{2} B_\nu\right)
- i \bar C_\mu \Delta^{(gh)\mu}{}_{\nu} C^\nu,
\label{gfghl}
\eea
where $a$ is another gauge parameter and
\bea
\Delta^{(gh)\mu}{}_{\nu} \equiv - \bg^{\mu\la} \bg^{\rho\s} \Big( \bnabla_\rho(g_{\la\nu} \nabla_\s
+ g_{\s\nu}\nabla_\la) - \frac{b+1}{d} \bnabla_\la(g_{\rho\nu}\nabla_\s+g_{\s\nu}\nabla_\rho) \Big).
\label{bfield}
\eea
Since the $B_\mu$ field involves no derivatives, if $a\neq 0$, we can simply integrate it out and
we are left with the gauge-fixing and FP ghost terms. However, this is not convenient for our purpose.

Under the rescaling transformation~\p{t1}, we have
\bea
\d \sqrt{\bg} = d\e \sqrt{\bg}, \qquad
\eea
where $\bg=\det(\bg_{\mu\nu})$. If we assume that the auxiliary field $B_\mu$ has dimension $d_B$,
the quadratic term in the auxiliary field transforms like
\bea
\d \Big(-\frac{a}{2} \sqrt{\bg} \bg^{\mu\nu} B_\mu B_\nu\Big)
= -\frac{a}{2} (d-2+2d_B) \e \sqrt{\bg} \bg^{\mu\nu} B_\mu B_\nu,
\eea
so this is invariant if we take
\bea
d_B = \frac{2-d}{2}, \quad
\mbox{or} \quad
\d B_\mu = \frac{2-d}{2}\e B_\mu.
\label{db}
\eea
The rescaling transformation of the gauge-fixing function is
\bea
\d F_\mu =-2\e F_\mu.
\label{tgff}
\eea
This leads to
\bea
\d (\sqrt{\bg} \bg^{\mu\nu} B_\mu F_\nu) = \frac{d-6}{2} (\sqrt{\bg} \bg^{\mu\nu} B_\mu F_\nu),
\label{gf}
\eea
so that this is invariant only for $d=6$, as found by Morris~\cite{Morris}.

Is there no way to make it invariant for any dimension? Actually if we take the Landau gauge $a=0$,
we do not have to take the rescaling dimension of the auxiliary field $B_\mu$ as \p{db}.
We can just require that \p{gf} is invariant to find
\bea
\d B_\mu=(4-d) \e B_\mu.
\label{bt}
\eea
Note that it is conceptually better to keep $B_\mu$ rather than to eliminate it for $a=0$.
According to the BRST symmetry~\p{brst1} and \p{brst2}, we should assign the rescaling dimension to each field as
\bea
\d C^\mu = 0, \qquad
\d \bar C_\mu = (4-d) \e \bar C_\mu.
\label{ct}
\eea
Note that $C^\mu$ and $\bar C_\mu$ are independent Hermitian fields, so they can have different dimensions.
It is then easy to see that the FP ghost term is also scale invariant.
We thus see that the Landau gauge is necessarily singled out by the rescaling invariance of these terms
in this formulation.
We note that Ref.~\cite{PV} achieved the invariance by introducing higher-derivative gauge fixing
without dimensionful parameters, and in that case we do not have to take the Landau gauge.
We thus confirm here in slightly improved way that higher-derivative gauge fixing is not necessary
if we choose the Landau gauge, as discussed in \cite{Morris}.

In the usual gauge $a\neq 0$, we have a contribution from the square term of the gauge-fixing function.
In the Landau gauge, there is also effectively the same contribution.
When we integrate over the auxiliary field $B_\mu$, this produces $\delta$-function,
which must be path integrated by the fields. Then we have to take into account the Jacobian from
the gauge-fixing function~\p{gff}. Because the Landau gauge strongly imposes the gauge condition to be zero,
the mode appearing there does not appear in other parts of the action. Denoting this mode by $\rho_\mu$,
it has a similar transformation property to the ghost term, and we can make it scale invariant
by assigning suitable rescaling dimension to that. We will also use the same cutoff for this term as
the FP ghost term. In this case we can see that the contribution almost cancels against the ghost term.
Since the discussion is basically the same as for the ghost, we suppress this
for the moment, and discuss this in more detail in sect~\ref{gftg}.
A related discussion is given in \cite{dsz2}.

Next come the cutoff terms that break the rescaling invariance.
Morris considered a cutoff that is related to the Hessians of the kinetic terms.
As a result, the consistency of the FRGE and the modified scale identity again requires
that the spacetime dimension be six.
However it was pointed out in \cite{PV} that this is not necessary if we use suitable cutoff.
Here we have to further modify that because the rescaling dimensions are different.
We thus consider
\bea
\Delta S_k (h_{\mu\nu}^T, \bg_{\mu\nu})
&=& \frac{1}{2}\int \sqrt{\bg} \left[h_{\mu\nu}^T \bg^{\mu\rho}\bg^{\nu\rho}R_k^T(\bar \Delta) h_{\rho\s}^T
+ h R_k(\bar \Delta) h \right], \nn
\Delta S_k^{gh} (\bar C_\mu, C^\mu, \bg_{\mu\nu})
&=& -i \int \sqrt{\bg}\, \bar C_\mu R_k^{gh}(\bar \Delta) C^\mu,
\eea
where we choose
\bea
R_k^T(\bar \Delta) = c k^{d-4} r(y), \qquad
R_k(\bar \Delta) = c_0 k^{d-4} r(y), \qquad
R_k^{gh}(\bar \Delta) = c_{gh} k^4 r(y),
\eea
with $y=\bar\Delta/k^2$ and suitable coefficients $c,c_0,$ and $c_{gh}$.
Here, $r$ is a dimensionless function that vanishes rapidly for $y>1$ and $r(0)=1$.
These are independent of any parameters in the action, and called a ``pure'' cutoff~\cite{NP}.
Note that the power of the cutoff $k$ is different from those in \cite{PV}.
This is necessary in order to achieve the nice transformation property of these terms.
Denoting $t= \ln k$, we then have
\bea
\d R_k^T = \e[\pa_t R_k^T-(d-4) R_k^T], \quad
\d R_k = \e[\pa_t R_k-(d-4) R_k], \quad
\d R_k^{gh} = \e[\pa_t R_k^T-4 R_k^{gh}],
\eea
for the above cutoffs, and
\bea
\d \Delta S_k
&=& \frac{\e}{2} \int \sqrt{\bg} \left[h_{\mu\nu}^T \bg^{\mu\rho}\bg^{\nu\rho} \pa_t R_k^T(\bar \Delta)
h_{\rho\s}^T + h \pa_t R_k(\bar \Delta) h -4d R_k(\bar \Delta) h \right], \nn
\d \Delta S_k^{gh}
&=& -i \e \int \sqrt{\bg}\, \bar C_\mu \pa_t R_k^{gh}(\bar \Delta) C^\mu.
\label{ftrans1}
\eea
Note that there is no term proportional to $\Delta S_k$ which was present in \cite{Morris} with
a factor $(d-6)$.

\subsection{Modified Ward identity}

The generating functional of the Green functions are given by
\bea
e^{W_k[\bg,J]} = \int [{\cal D}g] e^{-S[g]-\Delta S_k[\bg,h]+\int (J^{\mu\nu}h_{\mu\nu}^T+J_h h)} ,
\eea
where we have suppressed tensor indices. Since $W_k$ is a functional of $\bg_{\mu\nu}$ and $J$,
it does not have transformation under the variation of $h_{\mu\nu}$.
Therefore under the rescaling transformation with $J$ fixed, we get
\bea
\int \frac{\d W_k}{\d \bg_{\mu\nu}} \d \bg_{\mu\nu}
= - \lan \d\Delta_k S \ran + \int J^{\mu\nu} \lan \d h_{\mu\nu}^T \ran + \int J_h \lan \d h \ran.
\label{gg}
\eea
We now define
\bea
\G_k = -W_k +\int J^{\mu\nu} \lan h_{\mu\nu}^T \ran +\int J_h \lan h \ran
-\Delta S_k[\bg_{\mu\nu},\lan h_{\mu\nu}^T \ran, \lan h\ran],
\label{effectiveaction}
\eea
where $\lan h \ran$ stands for the expectation values of $h_{\mu\nu}^T$ and $h$.
We then note that
\bea
\frac{\d \G_k}{\d \bg_{\mu\nu}}
&=& - \frac{\d W_k}{\d \bg_{\mu\nu}}
 - \frac{\d \Delta S_k[\bg,\lan h_{\mu\nu}^T \ran, \lan h\ran]}{\d \bg_{\mu\nu}}, \nn
\frac{\d \G_k}{\d \lan h_{\mu\nu}^T\ran}
&=& J^{\mu\nu} - \frac{\d \Delta S_k[\bg,\lan h_{\mu\nu}^T \ran, \lan h\ran]}{\d \lan h_{\mu\nu}^T\ran}, \nn
\frac{\d \G_k}{\d \lan h \ran}
&=& J_h - \frac{\d \Delta S_k[\bg,\lan h_{\mu\nu}^T \ran, \lan h\ran]}{\d \lan h \ran}, \nn
\frac{\d W_k}{\d J^{\mu\nu}}
&=& \lan h_{\mu\nu}^T\ran, \qquad
\frac{\d W_k}{\d J_h}
= \lan h\ran.
\eea
Using these in Eq.~\p{gg}, we find~\cite{Safari}
\bea
\int\!\!\left( \frac{\d \G_k}{\d \bg_{\mu\nu}} + \frac{\d\Delta S_k}{\d \bg_{\mu\nu}} \right) \d\bg_{\mu\nu}
= \lan\d \Delta S_k\ran - \!\int\!\!\left( \frac{\d \G_k}{\d \lan h_{\mu\nu}^T \ran}
+ \frac{\d\Delta S_k}{\d \lan h_{\mu\nu}^T \ran} \right) \lan \d h_{\mu\nu}^T \ran
- \!\int\!\!\left( \frac{\d \G_k}{\d \lan h \ran}+ \frac{\d\Delta S_k}{\d \lan h \ran} \right) \lan \d h \ran.
\nn
\eea
Together with \p{ftrans1}, this leads to
\bea
\int \frac{\d \G_k}{\d \bg_{\mu\nu}} \d\bg_{\mu\nu} + \int\frac{\d\G_k}{\d \lan h \ran} \lan \d h \ran
&=& \frac{\e}{2} \int \sqrt{\bg} \bg^{\mu\rho} \bg^{\nu\s} \pa_t R_k^T \frac{\d^2 W_k}{\d J_{\mu\nu}\d J_{\rho\s}}
+ \frac{\e}{2} \int \sqrt{\bg} \pa_t R_k \frac{\d^2 W_k}{\d J_h\d J_h} \nn
&& \hs{20} -\, i \e \int \sqrt{\bg} \bar C_\mu \pa_t R_k C^\mu.
\eea
In this way we finally arrive at the mWI
\bea
&& \e \left[ 2\int \bg_{\mu\nu} \frac{\d \G_k}{\d \bg_{\mu\nu}} 
-2d \int \frac{\d \G_k}{\d \lan h\ran} 
-2 \int \lan h\ran \frac{\d \G_k}{\d \lan h\ran} \right] \nn
&& = \e\left[ \frac{1}{2} \Tr \Big\{ \Big( \frac{\d^2 \G_k}{\d h^T \d h^T}
 + R_k^T\Big)^{-1} \pa_t R_k^T \Big\}
+ \frac{1}{2} \Tr \Big\{ \Big( \frac{\d^2 \G_k}{\d h\d h} +R_k\Big)^{-1} \pa_t R_k \Big\}
\right. \nn && \hs{20} \left.
- \Tr \Big\{ \Big( \frac{\d^2 \G_k}{\d \bar C \d C} +R_k^{gh}\Big)^{-1} \pa_t R_k^{gh}\Big\} \right].
\eea
Apart from the factor $\e$, the RHS is identical to the RHS of the exact RG equation.
We thus get
\bea
\int \left[ 2 \bg_{\mu\nu} \frac{\d \G_k}{\d \bg_{\mu\nu}} 
-2d \frac{\d \G_k}{\d \lan h\ran} 
-2 \lan h\ran \frac{\d \G_k}{\d \lan h\ran} \right]
-k \frac{d\G_k}{dk} =0.
\label{mwi}
\eea

\subsection{Single-metric approximation}

We now make the standard single-metric approximation~\cite{Reuter}.
In this approximation, we keep only the dependence on the constant part $\bar h$ defined by
\bea
\bar h = \frac{1}{V}\int \sqrt{\bg} h, \qquad
h^\perp = h - \frac{1}{V}\int \sqrt{\bg} h,
\eea
where $V= \int\sqrt{g}$.
As discussed in detail by Morris~\cite{Morris}, we have
\bea
&& \int \frac{\pa\G}{\pa h} =\frac{\pa\G}{\pa \bar h}, \nn
&& \int h \frac{\pa\G}{\pa h} =\bar h \frac{\pa\G}{\pa \bar h} + \int h^\perp \frac{\pa\G}{\pa h^\perp}.
\eea
The last term in the second equation is discarded in our approximation.
Substituting these into \p{mwi}, we get
\bea
2 \int \bg_{\mu\nu} \frac{\d \G_k}{\d \bg_{\mu\nu}} 
-2d \frac{\d \G_k}{\d \bar h} - 2 \bar h \frac{\d \G_k}{\d \bar h}
-k \frac{d\G_k}{dk} =0.
\label{mwi2}
\eea
The solution is given by
\bea
\G = \hat \G_{\hat k}[\hat g_{\mu\nu}],
\eea
where $\hat g_{\mu\nu}$ is defined as
\bea
\hat g_{\mu\nu}=\left(1+\frac{\bar h}{d}\right) \bg_{\mu\nu},
\label{hg}
\eea
and
\bea
\hat k = k/\sqrt{1+\bar h/d}.
\eea
Thus we arrive at the same flow equation in the single-metric approximation with the background metric
$\bg_{\mu\nu}$ and cutoff scale $k$ replaced by $\hat g_{\mu\nu}$ and $\hat k$, respectively,
in arbitrary dimensions.
The metric $\hat g_{\mu\nu}$ is no longer a fixed background but becomes dynamical through dependence on $\bar h$.
We 
should consider that the solution describes an ensemble of different scales.
More precisely, it describes an infinite ensemble of background spacetimes related by
the rescaling $\bg_{\mu\nu} \to \bg_{\mu\nu}/\a^2$, which is compensated by
$\bar h\to(\bar h+d)\a^2-d$ and $k\to k\a$. 
The variables $\hat g_{\mu\nu}$ and $\hat k$ are scale independent.

In this way we have been able to extend the scale-invariant formulation to arbitrary dimensions
without using higher-derivative gauge-fixing or exponential parametrization.
What is important is to adopt a suitable cutoff and Landau gauge, which enable us to make
the gauge-fixing and FP ghost terms rescaling invariant.

\section{Modified Ward identity in the exponential split}

In a practical discussion of the FRGE, it is quite often useful to use the exponential parametrization.
In this section, we derive the mWI in the exponential split~\p{exp} without higher-derivative gauge fixing.
The main difference from the linear split will be in the rescaling dimension of various terms.

\subsection{Scale transformation and the derivation}

Let us require background rescaling independence under
\bea
\d\bg_{\mu\nu} = 2\e \bg_{\mu\nu},
\label{et1}
\eea
together with
\bea
\d h_{\mu\nu} = 2\e (h_{\mu\nu}-\bg_{\mu\nu}),
\label{et2}
\eea
so that the total metric~\p{exp} does not change~\cite{PV}.
Here we also decompose the fluctuation field into traceless and trace parts just as in \p{dec}.
Using \p{et1} and \p{et2}, and taking the trace, we find
\bea
\d h_{\mu\nu}^T = 2\e h^T_{\mu\nu}, \qquad
\d h =-2d\e,
\label{et3}
\eea
in contrast to \p{t5}.

The gauge-fixing and FP terms may be obtained as in the preceding section.
The BRST transformation of the fluctuation $h_{\mu\nu}$ is complicated and we adopt
the result in \cite{PV}. Denoting the metric split~\p{exp} as
\bea
\bm g=\bar{\bm g}e^{\bm X}, \qquad
\bm X= \bar{\bm g}^{-1} \bm h,
\eea
in a matrix notation, the BRST transformation corresponding to reparametrization is given by
\bea
\d_B {\bm X}= \d\lambda\frac{ad_{\bm X}}{e^{ad_{\bm X}}-\bm 1}(\bar{\bm g}^{-1} {\cal L}_C \bar{\bm g}
+{\cal L}_C e^{\bm X} e^{-\bm X}),
\label{brst3}
\eea
where $ad_{\bm X} \bm Y=[\bm X, \bm Y]$ and
\bea
{\cal L}_C g_{\mu\nu} = g_{\rho\nu} \nabla_\mu C^\rho + g_{\rho\mu}\nabla_\nu C^\rho.
\eea

We use the same gauge-fixing function~\p{gff} as in the previous section.
The gauge-fixing and FP terms are then given as~\cite{opp}
\bea
{\cal L}_{GF+FP}/\sqrt{\bg} \hs{-2}&=&\hs{-2}
i \d_B \left[\bg^{\mu\nu} \bar C_\mu \left(F_\nu+\frac{a}{2} B_\nu\right)\right]/\d\la \nn
\hs{-2} &=&\hs{-2} - \bg^{\mu\nu} B_\mu \left( F_\nu +\frac{a}{2} B_\nu\right)
- i \bar C_\mu \bg^{\mu\rho} \Delta^{(gh)}_{\rho\nu} C^\nu.
\label{gfghl2}
\eea
The full form of the FP ghost term can be found in \cite{PV} and is very complicated,
but in the single-metric approximation, it is much simpler and is given by~\cite{opp}
\bea
\Delta^{(gh)}_{\mu\nu} \equiv \bg_{\mu\nu} \bg^{\rho\s}\bnabla_\rho \bnabla_\s +
\Big(1-2 \frac{b+1}{d}\Big) \bnabla_\mu \bnabla_\nu +\br_{\mu\nu}.
\eea
What is important is that under the rescaling transformation~\p{et1} and \p{et2}, we have
\bea
\d \sqrt{\bg} = d\e \sqrt{\bg}, \qquad
\d F_\mu = 0, \qquad
\d(\bg^{\mu\rho} \Delta^{(gh)}_{\rho\nu})= -2\e \bg^{\mu\rho} \Delta^{(gh)}_{\rho\nu}.
\eea
Here we again encounter the same problem with the linear split unless $d=2$ or we take the Landau gauge.
In this gauge, we can just assign the rescaling properties as
\bea
\d B_\mu = (2-d) B_\mu, \qquad
\d \bar C_\mu = (2-d) \bar C_\mu, \qquad
\d C^\mu = 0,
\eea
so that the gauge-fixing and FP terms are invariant.
We again emphasize that we do not have to introduce higher-derivative gauge fixing
if we just take the Landau gauge.

For the cutoff terms, we adopt a cutoff that is similar to that used
by Percacci and Vacca~\cite{PV} but suitably modified in the ghost part.
Thus we consider
\bea
\Delta S_k (h_{\mu\nu}^T, \bg_{\mu\nu})
&=& \frac{1}{2}\int \sqrt{\bg} \left[h_{\mu\nu}^T \bg^{\mu\rho}\bg^{\nu\rho}R_k^T(\bar \Delta) h_{\rho\s}^T
+ h R_k(\bar \Delta) h \right], \nn
\Delta S_k^{gh} (\bar C_\mu, C^\mu, \bg_{\mu\nu})
&=& -i \int \sqrt{\bg}\, \bar C_\mu R_k^{gh}(\bar \Delta) C^\mu,
\eea
where we choose
\bea
R_k^T(\bar \Delta) = c k^{d} r(y), \qquad
R_k(\bar \Delta) = c_0 k^{d} r(y), \qquad
R_k^{gh}(\bar \Delta) = c_{gh} k^2 r(y).
\eea
Denoting $t= \ln k$, we then have
\bea
\d R_k^T = \e[\pa_t R_k^T-d R_k^T], \quad
\d R_k = \e[\pa_t R_k-d R_k], \quad
\d R_k^{gh} = \e[\pa_t R_k^T-2 R_k^{gh}],
\eea
for the above cutoffs, and
\bea
\d \Delta S_k
&=& \frac{\e}{2} \int \sqrt{\bg} \left[h_{\mu\nu}^T \bg^{\mu\rho}\bg^{\nu\rho} \pa_t R_k^T(\bar \Delta)
h_{\rho\s}^T + h \pa_t R_k(\bar \Delta) h -4d R_k(\bar \Delta) h \right], \nn
\d \Delta S_k^{gh}
&=& -i \e \int \sqrt{\bg}\, \bar C_\mu \pa_t R_k^{gh}(\bar \Delta) C^\mu.
\label{ftrans2}
\eea

Repeating the same manipulations as before, we find the mWI
\bea
&& 2\int \bg_{\mu\nu} \frac{\d \G_k}{\d \bg_{\mu\nu}} 
+2 \int \lan h^T_{\mu\nu} \ran \frac{\d \G_k}{\d \lan h^T_{\mu\nu}\ran}
-2d \int \frac{\d \G_k}{\d \lan h\ran} 
\nn
&& = \frac{1}{2} \Tr \Big\{ \Big( \frac{\d^2 \G_k}{\d h^T \d h^T}
 + R_k^T\Big)^{-1} \pa_t R_k^T \Big\}
+ \frac{1}{2} \Tr \Big\{ \Big( \frac{\d^2 \G_k}{\d h\d h} +R_k\Big)^{-1} \pa_t R_k \Big\}
\nn && \hs{20}
- \Tr \Big\{ \Big( \frac{\d^2 \G_k}{\d \bar C \d C} +R_k^{gh}\Big)^{-1} \pa_t R_k^{gh}\Big\} .
\eea
The main difference comes from the fact that here the traceless mode $h^T_{\mu\nu}$ transforms homogeneously
but the trace part $h$ inhomogeneously (see \p{et3}) in contrast to the linear split,
where $\d h^T_{\mu\nu}=0$ and $\d h=-2\e(h+d)$.

The right-hand side is identical to the right-hand side of the exact RG equation.
We thus get
\bea
\int \left[ 2 \bg_{\mu\nu} \frac{\d \G_k}{\d \bg_{\mu\nu}} 
+2 \int \lan h^T_{\mu\nu} \ran \frac{\d \G_k}{\d \lan h^T_{\mu\nu}\ran}
-2d \int \frac{\d \G_k}{\d \lan h\ran} 
 \right]
-k \frac{d\G_k}{dk} =0.
\eea

In the standard single-metric approximation~\cite{Reuter} but slightly extended here,
we keep only the dependence on the constant part $\bar h$.
We then find that the solution to the mWI is given by
\bea
\G = \hat \G_{\hat k}[\hat g_{\mu\nu}],
\eea
where $\hat k$ and $\hat g_{\mu\nu}$ are defined as
\bea
\hat k = e^{-\bar h/(2d)} k, \qquad
\hat g_{\mu\nu}= e^{\bar h/d} \bg_{\mu\nu}, \qquad
\hat h_{\mu\nu}^T = e^{\bar h/d} h_{\mu\nu}.
\label{hg2}
\eea
We again see that the solution to the FRGE is written in terms of scale-independent variables
and the coarse-graining problem may be resolved.

\subsection{Gauge-fixing term and ghost}
\label{gftg}

Here we discuss in more detail the contribution of the gauge-fixing term and FP ghost.
For this purpose, it is more convenient to use the York type decomposition
\bea
h_{\mu\nu} = h_{\mu\nu}^{TT} + \bnabla_\mu \rho_\nu +\bnabla_\nu \rho_\mu
-\frac{2}{d} \bg_{\mu\nu} \bnabla^\a \rho_\a + \frac{1}{d} \bg_{\mu\nu} h,
\label{yorkt}
\eea
with
\bea
\bnabla^\nu h_{\mu\nu}^{TT} = \bg^{\mu\nu} h_{\mu\nu}^{TT}=0.
\eea
We then find~\cite{dsz2}
\bea
F_\mu = \left[\bg_{\mu\nu} \bnabla^2 + \frac{d-2}{2d} (\bnabla_\mu \bnabla_\nu + \bnabla_\nu \bnabla_\mu)
+\frac{d+2}{2} \br_{\mu\nu} \right] \rho^\nu - \frac{b}{d} \nabla_\nu h,
\eea
and the Jacobian from this must be taken into account.
On the other hand, the ghost kinetic term is given by
\bea
\Delta^{(gh)}_{\mu\nu} = \bg_{\mu\nu} \bnabla^2 + \frac{d-2-2b}{2d} (\bnabla_\mu \bnabla_\nu
+ \bnabla_\nu \bnabla_\mu) +\frac{d+2+2b}{2} \br_{\mu\nu}.
\eea
We see that if we set $b=0$, these contributions would cancel each other~\cite{dsz2}.
For this, it is necessary to use the same cutoff as the ghost term.
In addition, in making the field transformation into $\rho_\mu$ in \p{yorkt},
we get the Jacobian $\Det(\Delta -\frac{\br}{d})^{1/2}$ from that, and it should be taken into account
with a similar cutoff as the FP ghost.
This is the spin-1 contribution discussed in \cite{opv,opv2}.

\section{Explicit example of the scale-independent flow equation for $f(R)$ gravity in $d=4$}

To get some idea of how the FRGE looks in this setting, let us consider an explicit example
of the theory known as $f(R)$ gravity whose action is
\bea
S=\int d^d x \sqrt{-g} f(R).
\eea
We study this system with the exponential parametrization~\p{exp}.
Since the preceding discussions show that the scale-invariant FRGE may be obtained just by replacing
the metric and momentum cutoff by scale-independent variables, we just check what FRGE is obtained
with our gauge-fixing and cutoff scheme.

Following the standard procedure, we find the only difference from the known result
is in the cutoff. We then arrive at the FRGE~\cite{opv,opv2}
\bea
\dot \G_k \hs{-1}&=&\hs{-1}\frac{1}{2} \mbox{Tr}_{(2)}
\left[\frac{\dot R_k^T(\Delta)}{f'(\br) \left( \Delta
+\alpha\br+\frac{2}{d(d-1)}\br \right)+ R_k^T(\Delta)}\right]
- \frac{1}{2} \mbox{Tr}_{(1)}\left[ \frac{\dot R_k^{gh}(\Delta)}{\Delta
+\gamma\br -\frac{1}{d}\br+ R_k^{gh}(\Delta)} \right]
\nn &&
+\; \frac{1}{2} \mbox{Tr}_{(0)} \left[ \frac{\dot R_k(\Delta)}
{f''(\br) \left( \Delta +\beta\br-\frac{1}{d-1}\br \right)
+\frac{d-2}{2(d-1)}f'(\br)+ R_k(\Delta)} \right],
\label{frge}
\eea
where the dot denotes the derivative with respect to the RG time $t= \log k/k_0$ (with $k_0$
an arbitrary reference scale)
and $\Delta =-\nabla^2$ is the Laplacian.
The subscripts on the traces represent contributions from different spin sectors:
$(2)$ denotes a trace over transverse-traceless symmetric tensor modes,
$(1)$ a trace over transverse-vector modes, and
$(0)$ a trace over scalar modes.
Here $\a$, $\b$, and $\c$ are free parameters, the choice of which corresponds to
the choice of RG schemes along with the choice of the function $R_k(z)$.
We note that the traces can in principle be evaluated for both negative and positive curvatures
and in any dimension $d$.
We give the necessary formulas in Appendix~\ref{heat}.

Evaluation of the traces is done as follows:
First, for some differential operator $z$, consider
\bea
\mbox{Tr}_{(j)}[W(z)] =\int_0^\infty ds \tilde W(s) \mbox{Tr}_{(j)} [e^{-sz}],
\label{step1}
\eea
for the spin-$j$ sector, where $\tilde W(s)$ is the inverse Laplace transform of $W(z)$:
\bea
W(z)=\int_0^\infty ds\,e^{-z s}\tilde W(s).
\eea
Using the heat kernel expansion
\bea
\mbox{Tr}_{(j)}[e^{-s z}]
= \frac{1}{(4\pi s)^{d/2}} \int_{S^d} d^d x \sqrt{\bg}\,
\sum_{n \geq 0} b_{2n}^{(j)} s^n \br^n,
\eea
in \p{step1}, we obtain
\bea
\mbox{Tr}_{(j)}[W(z)] =
\frac{1}{(4\pi)^{d/2}} \int_{S^d} d^d x \sqrt{\bg}\,
\sum_{n \geq 0} b_{2n}^{(j)} Q_{d/2-n}[W] \br^n,
\eea
where
\bea
Q_m[W] =\frac{1}{\G(m)} \int_0^\infty dz z^{m-1} W[z].
\eea

We choose the optimized cutoff profile \cite{optimized}
$r(y) = (1-y)\t(1-y)$, where $\theta$ is the Heaviside distribution.
For the contribution of the spin-2 modes in \p{frge}, we find
\bea
Q_m[W]_{(2)} = \frac{1}{\G(m)} \int_0^\infty dz z^{m-1}
\frac{(d-2)ck^{d-2}(k^2-z)+2ck^d}
{f'(\br)\left( z +\alpha\br+\frac{2}{d(d-1)}\br\right)+ck^{d-2}(k^2-z)} \t(k^2-z).
\eea
We use the dimensionless quantities $r =\br k^{-2}$,
$\vp(r) = k^{-d} f(\br)$,
$\dot f(\br)= k^d [ d \vp(r) - 2r \vp'(r) + \dot{\vp}(r) ]$,
$f'(\br)=k^{d-2} \vp'(r)$, and
$f''(\br)= k^{d-4} \vp''(r)$,
and define $\ta=\a+2/d(d-1)$.

From now on, we set $d=4$. We then obtain
\bea
\label{qtwo}
Q_m[W]_{(2)} \hs{-2}&=&\hs{-2} \frac{2c k^{2m}}{\G(m)} \int_0^1 dy y^{m-1}
\frac{2-y}{\vp'(r)\left( y+\ta r \right)+c(1-y)}
 \nn &=&\hs{-2}
\frac{2 c k^{2m}}{\G(m+2)[c+ \ta r\vp'(r)]^2} \Big[ 2(m+1)[c +\ta r\vp'(r)]
 \nn &&\hs{-2}
+m[c-\vp'(r)(2+\ta r)]\, _2F_1[1,1+m,2+m,\frac{c-\vp'(r)}{c+\ta r\vp'(r)}] \Big],~~~
\eea
where $_2F_1[a,b,c,z]$ is the hypergeometric function.
More explicitly
\bea
Q_2[W]_{(2)} \hs{-2}&=&\hs{-2} - \frac{c k^{4}}{(c-\vp'(r))^3}
\Big[ \{ c-\vp'(r)\}\{c-(3+2\ta r)\vp'(r)\}
 \nn &&\hs{-2}
+2[c+\ta r \vp'(r)][c-(2+\ta r)\vp'(r)] \log\Big\{\frac{(1+\ta r)\vp'(r)}{c+\ta r \vp'(r)}\Big\} \Big],
\nn
Q_1[W]_{(2)} \hs{-2}&=&\hs{-2} \frac{2c k^{2}}{(c-\vp'(r))^2}
\Big[c-\vp'(r)-[c-(2+\ta r)\vp'(r)]\log\Big\{\frac{(1+\ta r)\vp'(r)}{c+\ta r \vp'(r)}\Big\} \Big],
\nn
Q_0[W]_{(2)} \hs{-2}&=&\hs{-2} \frac{4c}{c+\ta r \vp'(r)}, \qquad
Q_{-1}[W]_{(2)} = -\frac{2c[c-(2+\ta r)\vp'(r)]}{k^2(c+\ta r \vp'(r))^2}.
\eea

Similarly, for spin 1 we find
\bea
\label{qone}
Q_m[W]_{(1)} = \frac{2 c^{gh}}{\G(m)} \int_0^\infty dz z^{m-1}
\frac{ k^2}
{z +\tc \br+c^{gh} (k^2-z)} \t(k^2-z),
\eea
where we have defined $\tc=\c-\frac14$. The relevant results are
\bea
Q_2[W]_{(1)} \hs{-2}&=&\hs{-2} \frac{2c^{gh} k^{4}}{(c^{gh}-1)^2}
\Big[1-c^{gh}-(c^{gh}+\tc r)\log\Big\{\frac{1+\tc r}{c^{gh}+\tc r}\Big\} \Big],
\nn
Q_1[W]_{(1)} \hs{-2}&=&\hs{-2} \frac{2c^{gh} k^{2}}{1-c^{gh}}
\log\Big\{\frac{1+\tc r}{c^{gh}+\tc r}\Big\} ,
\nn
Q_0[W]_{(1)} \hs{-2}&=&\hs{-2} \frac{2c^{gh}}{c^{gh}+\tc r},
\eea
while for spin 0 we have
\bea
\label{qzero}
Q_m[W]_{(0)}
&=& \frac{2 c_0 k^{2m}}{\G(m)} \int_0^1 dy y^{m-1} \frac{2-y}
{\vp''(r)\left( y +\tb r \right)+\frac{1}{3}\vp'(r) +c_0(1-y)},
\eea
with $\tb=\b-\frac13$.
We find
\bea
Q_2[W]_{(0)} \hs{-2}&=&\hs{-2} -\frac{c_0 k^{4}}{9[c_0-\vp''(r)]^3}
\Bigg[3 \{ c_0-\vp''(r) \} \{3c_0-2\vp'(r)-3(3+2\tb r)\vp''(r) \}
 \nn &&\hs{-2}
+2\Big[9 \{c_0+\tb r\vp''(r)\}\{c-(2+\tb r)\vp''(r)\}-\vp'(r)^2
 \nn &&\hs{-2}
\left. -6(1+\tb r)\vp'(r)\vp''(r)\Big]
\log\Big\{\frac{\vp'(r)+3(1+\tb r)\vp''(r)}{3c_0+\vp'(r)+3\tb r\vp''(r)}\Big\} \right],
\nn
Q_1[W]_{(0)} \hs{-2}&=&\hs{-2} \frac{2c_0 k^{2}}{3[c_0-\vp''(r)]^2}
\Bigg[3c_0-3\vp''(r)
 \nn &&\hs{-2}
-\{3c_0-\vp'(r)-3(2+\tb r)\vp''(r)\}
\log\Big\{\frac{\vp'(r)+3(1+\tb r)\vp''(r)}{3c_0+\vp'(r)+3\tb r\vp''(r)}\Big\} \Bigg],
\nn
Q_0[W]_{(0)} \hs{-2}&=&\hs{-2} \frac{12c_0}{3c_0+\vp'(r)+3\tb r \vp''(r)},
\nn
Q_{-1}[W]_{(0)} \hs{-2}&=&\hs{-2} \frac{6c_0\left[-3c_0+\vp'(r)+3(2+\tb r)\vp''(r)\right]}
{k^2[3c_0+\vp'(r)+3\tb r \vp''(r)]^2},
\eea

The heat kernel coefficients $b_{2n}$ for $\Delta$ acting on spin-2, 1, and 0 are given
in~\cite{opv,opv2} for our case and summarized in Appendix~\ref{heat}.
Substituting these heat kernel coefficients and Eqs.~\p{qtwo}, \p{qone}, and \p{qzero} in \p{frge}, we obtain
\bea
2(4\pi)^2 (\dot{\varphi} -2 r \varphi '+4 \varphi)
&=&
b_0^{(0)} q_2^{(0)} - b_0^{(1)} q_2^{(1)} + b_0^{(2)} q_2^{(2)}
\nn &&
+ \left( b_2^{(0)} q_1^{(0)} - b_2^{(1)} q_1^{(1)} + b_2^{(2)} q_1^{(2)}\right) r
\nn &&
+ \left( b_4^{(0)} q_0^{(0)} - b_4^{(1)} q_0^{(2)} + b_4^{(2)} q_0^{(2)}\right) r^2
\nn &&
+ \left( b_6^{(0)} q_{-1}^{(0)} + b_6^{(2)} q_{-1}^{(2)}\right) r^3,
\label{erge}
\eea
where we have defined $q_m^{(i)} \equiv Q_m[W]_{(i)}/k^{2m}$, which are $k$-independent.
The scale-independent solution is obtained from the solution of this flow equation just
by using the scale-independent variables~\p{hg2}.

We thus find the flow equation in terms of the scale-independent variables.
The explicit example looks rather complicated with logarithms.
The next problem would be to try to see what solutions it allows,
but a full analysis is beyond the scope of this paper.

\section{Conclusions}

In this paper we have been able to extend the scale-independent FRGE formulated in six dimensions~\cite{Morris}
to arbitrary dimensions.
The crucial point in achieving this is the recognition of the necessity of the Landau gauge and
a change of the cutoff scheme.
It has been pointed out that this can be also done if one uses the exponential split of the metric,
higher-derivative gauge fixing and pure cutoff~\cite{PV}. However, we believe that
the first condition, the exponential split, may not be necessary as we have shown.
It is certainly true that if we use higher-derivative gauge fixing, it is possible to realize the scale
independence for arbitrary gauge fixing parameters.
However the use of higher-order gauge fixing is unusual, and
it is nice to see that it is possible to formulate it with the often-used gauge fixing.
We have seen that it is indeed possible if we adopt the Landau gauge.
Thus the first two conditions, i.e. exponential split of the metric and higher-derivative gauge fixing
are not required, but the third is essential.

Even though this is true, the exponential split has various advantages like avoidance of
unphysical singularities~\cite{opv,opv2,FO,nink,GKL,opp}.
So we have studied the problem in both linear and exponential splits,
and have shown that it is possible to formulate the background independence in this approach.
These are not unique choices realizing the background independence, but most commonly used parametrizations.

To get some idea of what the resulting FRGE looks like, we have also given it for the case of $f(R)$ gravity.
Because we have to use the pure cutoff scheme, the resulting FRGE becomes quite complicated.
The full analysis of its solutions requires quite a lot of work, and is left for future study.
It would be very interesting to give solutions to this equation.

\section*{Acknowledgments}

I am grateful for numerous valuable discussions with Tim Morris and Roberto Percacci.
This work was supported in part by the Grant-in-Aid for
Scientific Research Fund of the Japan Society for the Promotion of Science (C) No. 16K05331.

\appendix
\section{Heat kernel coefficients on the $d$-sphere}
\label{heat}

The heat kernel coefficients can be found by summing over eigenvalues $\la_\ell(d,s)$ of
the operator $\Delta$ weighted by their multiplicity $M_\ell(d,s)$
\bea
\mbox{Tr}_{(s)}[ e^{-\s (\Delta+E_{(s)})}] =\sum_\ell M_\ell(d,s) e^{-\s(\la_\ell(d,s)+E_{(s)})}.
\eea
For general $d$, $\la_\ell(d,s)$ and $M_\ell(d,s)$ are summarized in Table~\ref{sphere}.
\begin{table}[h]
\begin{center}
\begin{tabular}{|c|c|c|c|}
\hline
Spin & Eigenvalue $\la_\ell(d,s)$ & Multiplicity $M_\ell(d,s)$ & \\
\hline
\hline
0 & $\frac{\ell(\ell+d-1)}{d(d-1)} \br$ & $\frac{(2\ell+d-1)(\ell+d-2)!}{\ell! (d-1)!}$
 & $\ell=0,1,\dots$ \\
\hline
1 & $\frac{\ell(\ell+d-1)-1}{d(d-1)} \br$
 & $\frac{\ell(\ell+d-1)(2\ell+d-1)(\ell+d-3)!}{(d-2)! (\ell+1)!}$
 & $\ell=1,2,\ldots$ \\
\hline
2 & $\frac{\ell(\ell+d-1)-2}{d(d-1)} \br$
 & $\frac{(d+1)(d-2)(\ell+d)(\ell-1)(2\ell+d-1)(\ell+d-3)!}{2(d-1)! (\ell+1)!}$
 & $\ell=2,3,\ldots$ \\
\hline
\end{tabular}
\end{center}
\caption{Eigenvalues and multiplicities of the Laplacian on the $d$-sphere.}
\label{sphere}
\end{table}

We use the Euler-MacLaurin formula
\bea
\sum_{n=a}^b f(n) = \int_a^b f(x) dx+\frac{f(b)+f(a)}{2}
+ \sum_{k=1}^\infty \frac{B_{2k}}{(2k)!} \left( f^{(2k-1)}(b)-f^{(2k-1)}(a) \right).
\label{EM}
\eea
Here $B_{2k}$ denotes the Bernoulli numbers, and the boundaries are
$a=2$ and $b=\infty$.
Naively one would expect that $a=0\; (s=0)$, $a=1\; (s=1)$, $a=2\; (s=2),$ and $b=\infty$.
However one has to leave out the mode $n=1$ for the spin-1 field $\xi_\mu$
(Killing vectors) and for the field $\sigma$ one has to leave out the modes
$n=0$ (constant) and $n=1$ (related to the five conformal Killing vectors
that are not Killing vectors), so the sum should start from $n=2$.

For $d=4$, the functions $f^{(s)}(x)$ entering
into \p{EM} are
\bea
f^{(0)}(x) \hs{-2}&=&\hs{-2} \frac16 (x+1)(x+2)(2x+3) e^{-\frac{1}{12}x(x+3) \br \s+ \b\br\s}, \nn
f^{(1)}(x) \hs{-2}&=&\hs{-2} \frac12 x(x+3)(2x+3) e^{-\frac{1}{12}\{x(x+3)-1\} \br \s+ \c\br\s}, \\
f^{(2)}(x) \hs{-2}&=&\hs{-2} \frac56 (x-1)(x+4)(2x+3) e^{-\frac{1}{12}\{x(x+3)-2\} \br\s+\a\br\s}.
\eea
The integral parts in \p{EM} are given by
\bea
\int_2^\infty dx f^{(0)}(x) \hs{-2}&=&\hs{-2}\frac{1}{(4\pi \s)^2} \int_{S^d} d^d x \sqrt{\bg}
 \left(1 + \br\s \right)e^{-\frac{5\br\s}{6} +\b \br\s}, \nn
\int_2^\infty dx f^{(1)}(x) \hs{-2}&=&\hs{-2}\frac{1}{(4\pi \s)^2} \int_{S^d} d^d x \sqrt{\bg}
 \left(3+\frac52 \br\s \right)e^{-\frac{3 \br\s}{4}+\c\br\s}, \nn
\int_2^\infty dx f^{(2)}(x) \hs{-2}&=&\hs{-2}\frac{1}{(4\pi \s)^2} \int_{S^d} d^d x \sqrt{\bg}
 \left(5+\frac{5}{2}\br\s \right)e^{-\frac{2\br\s}{3}+\a\br\s}.
\eea
We find the coefficients for $d=4$ given in Table~\ref{heatkernel}~\cite{opv,opv2}.
\begin{table}[h]
\begin{center}
\begin{tabular}{|c|c|c|c|c|}
\hline
Spin & $b_0$ & $b_2$ & $b_4$ & $b_6$ \\
\hline
0 & $1$ & $\frac16+\b$ & $\frac{-511+360\b+1080\b^2}{2160}$
& $\frac{19085-64386 \b+22680 \b^2+45360 \b^3}{272160}$ \\
\hline
1 & $3$ & $\frac14+3\c$ & $\frac{-607+360\c+2160\c^2}{1440}$
& $\frac{37259-152964 \c+45360\c^2+181440\c^3}{362880}$ \\
\hline
2 & $5$  & $-\frac56+5\a$ & $\frac{-1-360\a+1080\a^2}{432}$
& $\frac{311-126\a-22680\a^2+45360\a^3}{54432}$ \\
\hline
\end{tabular}
\end{center}
\caption{Heat kernel coefficients for $d=4$.}
\label{heatkernel}
\end{table}


\end{document}